**Maternal obesity alters the cerebrovasculature and clearance of β-amyloid in adult offspring**

Cheryl A. Hawkes[1], Victoria Goss[2], Elina Zotova[2], Tual Monfort[3], Anthony Postle[1,2], Sumeet Mahajan[3], James A.R. Nicoll[1], Roy O. Weller[1] and Roxana O. Carare[1]

[1]Clinical and Experimental Sciences, Faculty of Medicine, University of Southampton, South Lab and Pathology Block, Southampton General Hospital, Southampton, UK, SO16 6YD

[2]Southampton NIHR Respiratory BRU, Southampton Centre for Biomedical Research, Mass Spectrometry Unit, Southampton General Hospital, Southampton, UK, SO16 6YD

[3]Institute for Life Sciences and Department of Chemistry, Faculty of Natural and Environmental Sciences, University of Southampton, Southampton, UK, SO17 1BJ

Running title: Impact of maternal high fat on neurovascular unit of adult offspring

Corresponding author:

Dr. Cheryl Hawkes
Clinical and Experimental Sciences
South Lab and Pathology Block
LD66 (Mailpoint 806)
Southampton General Hospital
Tremona Road
Southampton, Hampshire
SO16 6YD, United Kingdom

Email: c.hawkes@soton.ac.uk
Phone: +44 (0)2380
Fax: +44 (0)2380 796085

Conflict of interest statement : The authors declare no conflict of interest.

**Abstract**

Maternal obesity is associated with increased risk of diabetes, cardiovascular disease and hypertension in adult offspring. Mid-life hypercholesterolemia and hypertension are risk factors for Alzheimer's disease, suggesting that the aging brain may be impacted by early life environment. We found that exposure to a high fat diet during gestation and lactation induced changes in multiple components of the neurovascular unit, including a downregulation in apolipoprotein E and fibronectin, an upregulation in markers of astrocytes and perivascular macrophages and altered blood vessel morphology in the brains of adult mice. Feeding of high fat diet after weaning increased lipid droplets in the brain and influenced the fatty acid composition of phosphatidylcholine and phosphatidylethanolamine species, but did not affect the neurovascular unit. Sustained high fat diet over the entire lifespan resulted in additional decreases in levels of pericytes and collagen IV, changes in phospholipid composition and impaired perivascular clearance of β-amyloid (Aβ) from the brain. In humans, vascular Aβ load was significantly increased in the brains of aged individuals with a history of hypercholesterolemia. These results support a critical role for early dietary influence on the brain vasculature across the lifespan, with consequences for the development of age-related cerebrovascular and neurodegenerative diseases.

**Introduction**

The global prevalence of obesity has risen sharply over the past 20 years, with approximately one third of US women of childbearing age currently estimated to be overweight or obese (1). In addition to increasing the risk of miscarriage, pre-eclampsia and thromboembolism, maternal obesity negatively impacts upon placental, embryonic and fetal growth (2). Obesity during gestation is linked with increased risk of coronary heart disease, diabetes, obesity and premature death in adult offspring (3, 4).

Obesity-related complications, including diabetes, hypertension and hypercholesterolemia, are risk factors for the development of sporadic Alzheimer's disease (AD), but the mechanisms that underscore this susceptibility are unknown (5). AD is characterized pathologically in part by the deposition of β-amyloid peptides (Aβ) in the parenchyma and in the vasculature as cerebral amyloid angiopathy (CAA). The hippocampus, entorhinal, parietal and occipital cortices are most often affected by Aβ deposition and CAA (6, 7). In the majority of AD cases, accumulation of Aβ is proposed to result from a failure of the mechanisms that degrade and remove Aβ from the brain (8). Components of the neurovascular unit, which includes astrocytes, pericytes, endothelial cells, smooth muscle cells and basement membranes (9), mediate the clearance of Aβ, and changes in the composition, structure, or activity of the neurovascular unit reduces the efficiency by which Aβ is removed from the brain (10). We have previously demonstrated that basement membranes are conduits by which Aβ is eliminated from the brain and that this mechanism fails with increasing age and in the presence of the APOE4 genotype (11-13).

Administration of high fat or high cholesterol diets in mouse models of AD has been shown to increase Aβ production (14, 15). However, less is known about the effects of high fat diet on the pathways that mediate Aβ clearance. The Latent Early-Life Associated Regulation (LEARn) hypothesis of AD postulates that exposure to environmental agents during early developmental stages induce epigenetic modifications that translate into pathological features later in life (16, 17). Based on the observations that 1) maternal obesity correlates with the development of metabolic diseases in adult offspring and 2) these diseases are risk factors for AD, we established a model of pre- and postnatal high fat feeding to test the hypothesis that the adult brain vasculature and clearance of Aβ is impacted by high fat diet exposure during development and early life.

We found that maternal obesity during gestation and lactation induced an upregulation in markers of astrocytes and perivascular macrophages and a downregulation of apolipoprotein E and fibronectin in the brains of adult mouse offspring. Feeding of high fat diet after weaning influenced the fatty acid composition of cerebral phospholipid species and increased lipid droplets in the brain, but did not affect the expression levels of cells within the neurovascular unit. Sustained high fat diet throughout life resulted in additional decreases in levels of pericyte markers and collagen IV and impaired perivascular clearance of Aβ from the brain. Vascular Aβ load was significantly increased in the brains of aged humans with a history of hypercholesterolemia. These results support a role for early environmental influence on the brain vasculature, with consequences for the development of cerebrovascular and neurodegenerative diseases in later life.

**Results**

*Phenotype of offspring exposed to pre-and/or postnatal high fat diet*

To establish a model of pre- and postnatal high fat feeding, dams were maintained on either a standard (C) or high fat (HF) diet before conception and during gestation and lactation.  At weaning, offspring were assigned either C or HF diet, generating four experimental groups: C/C, C/HF, HF/C, HF/HF, representing pre- and post-weaning diet, respectively (Fig. 1A) (18).  Dams fed the HF diet weighed significantly more during both the preconception and gestational period than those on the C diet ($p<0.0005$, Fig. 1B).  Both C/HF and HF/HF mice showed a significantly greater body weight versus the C/C and HF/C groups within 8 and 4-6 weeks post-weaning, respectively ($p<0.0005$; Fig. 1C).  At sacrifice, dams on the HF diet had a significantly higher plasma glucose level than control dams (Fig. 1D), but glucose levels did not differ between offspring groups (Fig. 1E), nor did brain weights (data not shown).   Assessment of liver histology revealed little lipid accumulation and a normal hepatic architecture in the C/C group, while mild steatosis was observed in the livers of the C/HF group (Fig. 1F).  A few Oil red O-stained lipid particles were observed in the livers of the HF/C group, whereas extensive steatosis and fat accumulation was seen in the HF/HF group (Fig. 1F).

*Effect of high fat diet on brain phospholipid profile, lipid droplets and markers of cholesterol transport*

Cerebral phospholipids are rich in polyunsaturated fatty acids (PUFAs) that cannot be manufactured *de novo* and are thus influenced by dietary intake (19-21). To determine if pre- and post-natal high fat diet affects fatty acid composition of two major classes of phospholipids, phosphatidylcholine (PC) and phosphatidylethanolamine (PE), membrane-free tissue fractions were isolated from the hippocampus and analyzed by mass spectrometry. Within the PC class, a modest but significant increase in PC 38:6 and PC 40:6 species was observed in C/HF mice compared to C/C ($p = 0.029$ and $p = 0.002$, respectively) and HF/HF mice ($p = 0.028$ and $p = 0.022$, respectively; Table 1). PC 38:4 was also significantly increased in the brains of HF/C mice compared to HF/HF mice ($p = 0.021$), while PC 38:5 was significantly decreased ($p = 0.007$; Table 1). Analysis of PE composition indicated that PE 38a:0, PE 40:3 and PE 42:2a species were increased by 16%, 32% and 20% respectively within HF/HF mice compared to the brains of HF/C animals ($p = 0.048$, $p = 0.042$, and $p = 0.008$, respectively; Table 1). By contrast, PE 38:4 was significantly decreased in HF/HF versus HF/C mice ($p = 0.007$; Table 1).

Fatty acids and sterols can be taken up by cells and stored as neutral lipids (e.g. triacylglycerols and cholesterol esters) in lipid droplets that function as both energy depots and protective bodies (22). To determine if lipid droplet accumulation in the brain was influence by pre- and post-natal exposure to high fat diet, hippocampal sections were analyzed using Coherent anti-Stokes Raman spectroscopy (CARS). A significantly greater hippocampal area was covered by lipid droplets in the brains of both C/HF and HF/C mice compared to mice in the C/C group (p = 0.003, Fig. 2A). Although the size of the lipid droplets appeared larger in the HF/HF group versus the C/C mice, total area covered by lipid deposits did not differ between the two groups (Fig. 2A).

Cerebral lipid homeostasis is regulated by the interaction between apolipoproteins (Apo), ATP-binding cassette transporters (ABC) and lipoprotein receptors including the low-density lipoprotein receptor related protein-1 (LRP-1). Possession of ApoE4 is a major genetic risk factor for AD (23), although ApoA-I and ABCA1 have also recently been implicated in the pathogenesis of CAA (24, 25). To determine if early life exposure to high levels of cholesterol affects the expression of proteins that regulate cholesterol transport in the brain, hippocampal samples were assessed for levels of ApoE, ApoA-I, ABCA1 and LRP-1. Western blot analyses revealed that pre-natal high fat exposure either alone or in combination with high fat diet post-weaning resulted in a significant downregulation of hippocampal ApoE levels compared to C/C and C/HF mice (p = 0.002; Fig. 2B). No differences were noted in the level of expression of LRP-1, ApoA-I or ABCA1 between dietary groups (Fig. 2C-E).

*Effect of high fat diet on components of the neurovascular unit*

Multiple components of the neurovascular unit function to clear Aβ from the brain (10, 12, 26). To determine the effect of pre- and post-natal HF diet on the neurovascular unit, the level of expression of the major proteins within the neurovascular unit was assessed.

Levels of platelet-derived growth factor receptor β (PDGFRβ), a marker of activated and proliferating pericytes, were significantly decreased in the hippocampi of HF/HF mice compared to the C/C group (p = 0.016; Fig. 3A). Conversely, the astrocytic glial fibrillary acidic protein (GFAP) was significantly upregulated in both the HF/C and the HF/HF group versus the C/C group and between HF/HF and C/HF mice (p = 0.001; Fig. 3B). Levels of the CD206 receptor expressed by perivascular macrophages were significantly upregulated in the hippocampi of HF/C and HF/HF mice compared to C/C and C/HF groups (p <0.0001; Fig. 3C). No differences were observed in the levels of the endothelial cell marker CD31 (Fig. 3D), the tight junction protein occludin (Fig. 3E) or α-smooth muscle actin (Fig. 3F) between any of the offspring groups.

Assessment of cerebrovascular basement membrane proteins showed that levels of collagen IV were significantly decreased in HF/HF mice compared to all other dietary groups (p = 0.010; Fig. 4A). Fibronectin levels were significantly lower in HF/C mice versus C/C (p = 0.021) and C/HF groups (p = 0.004; Fig. 4B). No differences were observed in levels of laminin, nidogen 2 and perlecan (Fig. 4C-E). Morphologic analysis of collagen IV and fibronectin showed a smooth distribution along hippocampal arteries in the C/C group, with a slightly more rugged appearance in the C/HF mice (Fig. 5A and B). In the HF/C mice, vessels stained with collagen IV had a significantly reduced diameter compared to all other dietary groups (p<0.0005, Fig. 5A and C). Fibronectin expression appeared reduced in the HF/C group compared to C/C and C/HF brains, but vessel diameter was not different (Fig. 5A and C). Less staining of both collagen IV and fibronectin was also noted in the HF/HF brains (Fig. 5A and B).

*Effect of high fat diet on perivascular drainage of Aβ from the brain*

We have previously shown that alterations to basement membrane protein expression affect the efficiency of Aβ drainage from the brain (11-13). To assess the pattern of perivascular drainage of Aβ in the four dietary groups, mice were injected intra-hippocampally with fluorescently-conjugated human Aβ40. In C/C and C/HF mice, Aβ distributed in an even pattern along laminin-positive basement membranes in the tunica media of hippocampal arteries (Fig. 6A). In HF/C mice, Aβ had a smooth distribution along some portions of the basement membrane but followed the ruffled appearance of the basement membrane in other portions of the artery (Fig. 6A). The pattern of drainage of Aβ in the arteries of HF/HF mice was significantly disrupted, with Aβ observed as large deposits within arterial walls (Fig. 6A).

ApoE co-localizes with fibrillar Aβ within perivascular spaces in human AD brains (27) and has been suggested to transport Aβ across the blood-brain barrier (28). To determine if soluble Aβ is associated with ApoE within basement membranes, brains from mice from each dietary group were processed for double labeling immunocytochemistry with antibodies against GFAP and ApoE. As expected, a notable decrease in the expression of ApoE was observed in the brains of HF/C and HF/HF mice compared to C/C and C/HF mice (Fig. 6B). However, although ApoE was expressed by perivascular astrocytes, no co-localization was seen between Aβ and ApoE, regardless of the pattern of drainage of Aβ (Fig. 6B).

We further observed that the intensity of the fluorescent signal of Aβ around the injection site appeared stronger in the hippocampi of HF/C mice, suggesting that diffusion of Aβ within the interstitial space may differ between dietary groups. To address this possibility, the intensity of Aβ-related fluorescence in the hippocampus relative to background fluorescence was quantified at the site of injection and in brain sections 400 μm posterior to the injection site. The mean fluorescence intensity at the injection site was significantly higher in the hippocampi of HF/C mice compared to the other dietary groups (p = 0.020; Fig. 6C). However, this difference in intensity was not observed away from the injection site (Fig. 6C), suggesting that more Aβ was retained at the injection site in the brains of HF/C mice.

*Effect of hypercholesterolemia on basement membrane proteins and Aβ deposition in human brains*

Mid-life hypercholesterolemia is a risk factor for AD (29). To determine if basement membrane protein expression was also affected in the brains of humans with a history of hypercholesterolemia, tissue sections from the occipital cortex, a region of high CAA pathology (7), of individuals with normal and high levels of cholesterol were assessed for the expression of collagen IV and fibronectin. There was a significant reduction in cortical coverage by small diameter, collagen IV-positive blood vessels in individuals with hypercholesterolemia (p = 0.013, Fig. 7A). No differences were noted between groups in the percent cortical coverage by large diameter vessels or in total collagen IV-positive vessels (Fig. 7A). Fibronectin expression did not differ between normal and high cholesterol groups when small and large diameter vessels were analyzed separately, but was significantly upregulated in the high cholesterol group when the cortical coverage of fibronectin staining for all vessels was calculated ($p = 0.043$; Fig. 7B).

To determine the effect of hypercholesterolemia on Aβ accumulation, cortical sections were assessed for Aβ deposition. Individuals with high levels of cholesterol had a significantly greater CAA load than those with normal cholesterol ($p = 0.007$; Fig. 7C and D). Cholesterol levels did not impact upon the percent of cortical coverage by parenchymal plaques or total Aβ deposition (Fig. 7C and D).

**Discussion**

In the present study, we found that exposure to high fat during brain development affects the expression of markers of astrocytes, perivascular macrophages and basement membrane proteins (Table 2). High fat feeding throughout both the pre- and postnatal period further alters activity of pericytes and impairs the efficiency of Aβ clearance from the brain along cerebral basement membranes. Feeding of high fat diet after weaning influenced the fatty acid composition of PC and PE species, and increased the number of lipid droplets in the hippocampus, but did not affect the neurovascular unit or perivascular drainage of Aβ. These results support a role for early life environmental influence on the brain not just into adulthood, but across the lifespan, with consequences for the development of age-related cerebrovascular and neurodegenerative diseases.

There is a well established correlation between maternal obesity and the long-term health of the offspring, including increased risk of coronary heart disease, diabetes, hypertension and stroke in adulthood (30, 31). However, the effect of maternal high fat feeding on offspring brain development has been less well studied. Altered hypothalamic responses have been noted in the brains of murine offspring born to obese mothers, making them more likely to increase food intake and activate reward centers in response to fatty food consumption (32-34). Intake of high fat or high cholesterol diet influences lipid metabolism and phospholipid composition of the rodent brain (20, 21). Of the PC and PE fatty acids that we observed to be altered in the C/HF and HF/HF groups, the majority contained a PUFA, supporting the influence of postnatal diet on fatty acid availability in the brain (35). However, this effect was also influenced by dietary background, as the majority of the differences in PC fatty acid composition were observed in the C/HF group, while in the HF/HF group, the PE species were predominantly affected. These data suggest that prenatal diet may also influence the degree to which levels of PE and PC decrease in the AD brain (36-39). We also found that the hippocampal area covered by lipid droplets was increased in the brains of C/HF and HF/C but not in the HF/HF group compared to C/C mice. Lipid droplets are composed principally of neutral lipids, including triacylglycerols and cholesterol esters, which are bound by a phospholipid monolayer (22, 40). These organelles can act as energy reservoirs and can also remove potentially harmful excess lipid species (22, 41, 42). Thus, the observed increase in lipid droplets in the C/HF and HF/C groups may reflect a compensatory reaction by the brain to a novel high fat diet exposure. That this increase is not observed in the HF/HF mice suggests that sustained exposure to a high fat diet alters lipid storage mechanisms, and may have consequences for the production and clearance of Aβ (43).

The importance of cerebrovascular health in the pathogenesis of AD is increasingly recognized. Factors that affect vascular function, including diabetes, hypertension, hypercholesterolemia and apoE4 genotype all increase the risk of developing AD (44). Multiple components of the neurovascular unit and associated cells, including endothelial and smooth muscle cells, pericytes, perivascular macrophages, astrocytes and arterial basement membranes participate in the clearance Aβ from the brain (9, 11, 12, 45). Our findings that prenatal high fat diet exposure affected the activity and/or number of astrocytes, perivascular macrophages and basement membrane proteins in the adult brain, suggest that the neurovascular unit is influenced by early life dietary environment and that this impact is maintained into adulthood. Altered activity or efficiency of the neurovascular unit has previously been shown to reduce the efficiency by which Aβ is removed from the brain (10, 12, 13). We found that vessel morphology was dramatically affected in the brains of mice exposed to high fat during gestation and lactation and that diffusion of Aβ within the extracellular space and its clearance along arterial basement membranes was impaired in these mice. Interestingly, in the brains of HF/HF mice, Aβ appeared as discrete, deposit-like structures within the vessel walls, suggesting that perivascular drainage was also disrupted in the brains of these mice, but in a different pattern to that observed in the HF/C mice. Whether this is due to the additional alterations in markers of pericytes and basement membranes in the brains of the HF/HF mice or to other factors remains to be determined.

We have previously reported that perivascular drainage of Aβ along basement membranes is impaired in the presence of ApoE4 (13). Binding of Aβ to ApoE has been proposed as a mechanism by which Aβ is transported across the blood-brain barrier (28) and levels of ApoE are lower in ApoE4-positive individuals than in ApoE3 carriers (46). However, recent work has demonstrated minimal direct physical interaction between ApoE and Aβ under physiologic concentrations (47). We did not observe colocalization between soluble Aβ and ApoE under conditions of either normal (e.g. in C/C and C/HF mice) or disrupted (e.g. HF/C and HF/HF mice) perivascular drainage. Further, we observed different patterns of drainage in the brains of HF/C and HF/HF mice, despite significantly reduced levels of ApoE in both treatment groups. These data suggest that perivascular drainage of soluble Aβ is not mediated by a physical interaction with ApoE.

Age-related changes in the cerebrovasculature have been previously reported in the mouse brain, as well as human aged and AD brains (12, 48-50). We found that aged humans with a history of hypercholesterolemia had altered expression of fibronectin and collagen IV and increased amounts of vascular Aβ compared to aged individuals with normal cholesterol levels. Although the gestational and early life diet of these individuals is unknown, these data suggest a negative influence of high cholesterol on vascular basement membranes and their ability to clear Aβ from the aging human brain. Recent findings showing an association between slow prenatal growth and lower cognitive ability in late life (51), support a role for the early life environment in the aged human brain.

Collectively, these data indicate a critical role for prenatal diet in the development of the neurovascular unit and the efficiency of the adult cerebrovasculature to remove Aβ from the brain. These findings highlight maternal and gestational health as an important target for the prevention of neurodegenerative diseases.

**Materials and Methods**

*Animal Model:* Female C57BL6J mice were maintained on either a C (21% kcal fat, 17% kcal protein, 63% kcal carbohydrate, n=23) or HF chow diet (45% kcal fat, 20% kcal protein, 35% kcal carbohydrate; Special Diet Services, UK, n=44), as previously described(18). Dams were fed the HF diet four weeks before conception and during gestation and lactation. Dams were weighed once per week until mating, then at bi-weekly intervals until parturition. Litter size was standardized to a maximum of six pups to ensure no litter was nutritionally biased. At weaning, male offspring were assigned either C or HF diet, generating four experimental groups: C/C (n=27), C/HF (n=27), HF/C (n=27), HF/HF (n=27). Offspring body weights were recorded at two-week intervals up to 15 weeks post-weaning, at which time offspring were killed. All animals were allowed food and water *ad libitum*. All experiments received approval by the University of Southampton animal care committee and the Home Office (PPL 30/3095).

*Plasma assays:* Mice (n=5/group) were fasted overnight, whole blood collected and spun down to collect plasma. Plasma glucose levels were assessed using a commercially available kit (Abcam, Cambridge, UK).

*Liver Histology:* Mice (n=5/group) were perfused with 0.01M phosphate buffered saline (PBS) alone or followed by 4% paraformaldehyde. Liver sections were stained with hematoxylin-eosin and Oil Red O for assessment of steatosis and lipid accumulation, respectively.

*Brain phospholipid analysis:* Hippocampi (n=5/group) from perfused mice were sonicated in 0.01M PBS containing 20 μL of butylated hydroxytoluene (20g/L in methanol), centrifuged (5,700g, 5 mins) and supernatants collected, mixed with 0.6g potassium bromide and centrifuged again (287,000g, 2 hrs, 16°C). Total lipid was extracted using dichloromethane and methanol, as adapted from (52). Extracts were introduced by direct infusion into a triple quadrupole mass spectrometer (Xevo TQ MS, Waters, UK) equipped with an electrospray ionization interface. Data were processed using MassLynx software (Waters) and analyzed using a custom-designed macro (53).

*Coherent anti-Stokes Raman spectroscopy:* Mice (n=5/group) were perfused with PBS, followed by 4% paraformaldehyde and brain tissue sections were cut on a cryostat (20 μm thickness) and collected onto microscope slides. A home-built CARS setup comprising of a Chameleon (Coherent) and Compact OPO (APE Berlin) coupled to an inverted Nikon Ti-U 2000 microscope was used to acquire images. The beams were temporally overlapped using a delay stage and combined to form a spatially overlapped collinear beam. The pump beam was set to 835 nm the Stokes beam from the OPO was tuned to 1195 nm to target the $CH_2$ stretching Raman frequency of 2850 cm$^{-1}$. At least 4 images were taken in the hippocampal regions of each brain section with a 20x (NA: 0.75) objective. Two brain sections were imaged per animal. Dwell times of 30 μs were usually chosen and an area of 120 μm x 120 μm scanned at 512 x 512 pixels. The highly active lipid droplet areas were quantified using a code written in MATLAB (Mathworks, UK). For quantification the number of pixel with intensities exceeding a certain threshold was counted as described earlier (54).

*Western blotting*: Hippocampi (n=11/group, n=5-6/antibody) from perfused mice were processed and separated by gel electrophoresis as previously described (11). Membranes were incubated overnight at 4°C with specified primary antibodies (Table 3), stripped and re-probed with anti-glyceraldehyde-3-PDH (GAPDH) antibody to ensure equal protein loading. Immunoblots were quantified by densitometry using Image J software (NIH, Maryland, USA) and calculated as an optical density ratio of protein levels normalized to GAPDH levels.

*Immunocytochemistry:* Mice (n=5/group) were perfused with PBS, followed by 4% paraformaldehyde and processed for single- and double-labeling immunocytochemistry as previously described (11-13). Brain tissue sections were incubated with anti-collagen IV (1:500), anti-fibronectin (1:500), anti-$\alpha$ smooth muscle actin (1:500), anti-GFAP (1:1000) and anti-ApoE (1:250) and developed with nickel-enhanced diaminobenzidine as chromogen or with AlexaFluor-conjugated secondary antibodies (Invitrogen, Paisley, UK). Photomicrographs were captured using a Leica SP5 confocal laser scanning microscope (Milton Keys, UK) and exported to Photoshop CS software. Differences between mean arterial thicknesses were calculated using Image J software.

*Intracerebral injections:* Mice (n=6/group) were injected stereotaxically with 0.5 μL ice-cold HyLite Fluor 488-conjugated human A$\beta$40 (100 μM, AnaSpec, USA) into the left hippocampus, as described previously (11). Brains processed for double labeling immunocytochemistry for anti-laminin (1:500) and anti-$\alpha$ smooth muscle actin (1:500), as above. Quantification of fluorescence intensity of A$\beta$40 at the injection site and 400 μm posterior was calculated using Image J software by dividing the mean grey scale intensity within the ipsilateral hippocampus by the mean intensity of the overlying cortex.

***Human brain samples and immunocytochemistry:*** Sections of postmortem human brain from the occipital cortex were obtained from the Parkinson's UK Brain Bank, Imperial College London. Samples were collected and prepared in accordance with the National Research Ethics Service approved protocols. Cases were selected for normal (total cholesterol <5.0 mmol/L cholesterol, n = 30) or high cholesterol (total cholesterol >5.0 mmol/L, n = 25), and were matched for age (normal cholesterol mean age = 76.6±1.44 years; high cholesterol mean age = 76.9±1.34 years) and sex (normal cholesterol males n= 20, females n = 10; high cholesterol males n = 17, females n = 8). Brain tissue sections were processed as described previously (12) and incubated with anti-collagen IV (1:500) or anti-fibronectin (1:400) or anti-Aβ (1:100, clone 4G8, Covance, Leeds, UK). Micrographs were converted to binary images (4 images/section along one sulcus) and evaluated by densitometry using Image J software.

***Statistics:*** Data were checked for normality of distribution by Shapiro-Wilk test, F-test for equal variance and outliers identified using Grubb's test. Differences between means of two groups were calculating using two-sided Student's t test or Mann-Whitney U test when normality assumption was not met. For comparisons between multiple groups, one-way or two-way ANOVA with Bonferroni post hoc test was applied using SPSS. Significance was set at $p< 0.05$.

**Acknowledgements**

The authors wish to thank Dr. Felino Cagampang, University of Southampton for his help in setting up the mouse model of maternal obesity. SM would like to thank the EPSRC

Laser Loan Pool for providing the UFL3 system. This work was supported by Alzheimer's Research UK, the Wessel Medical Trust and the Faculty of Medicine, University of Southampton.

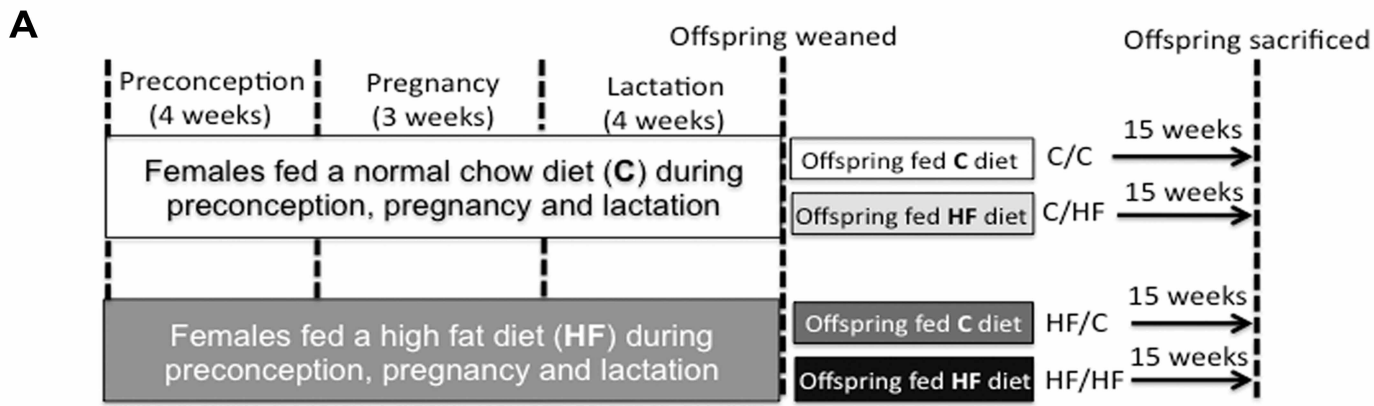
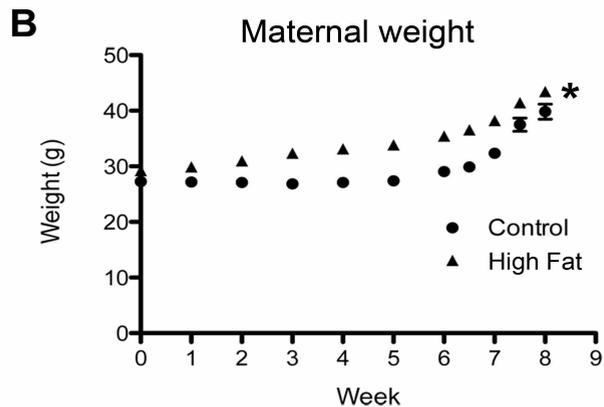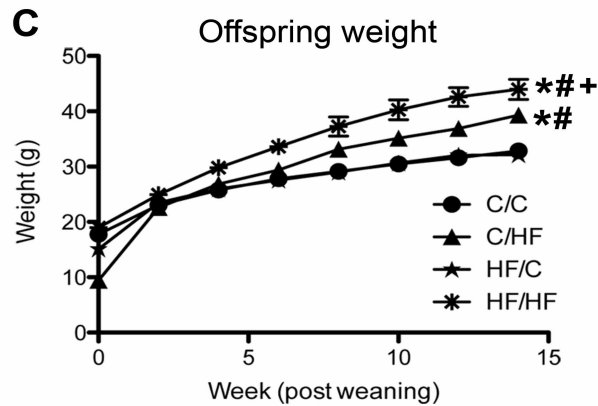
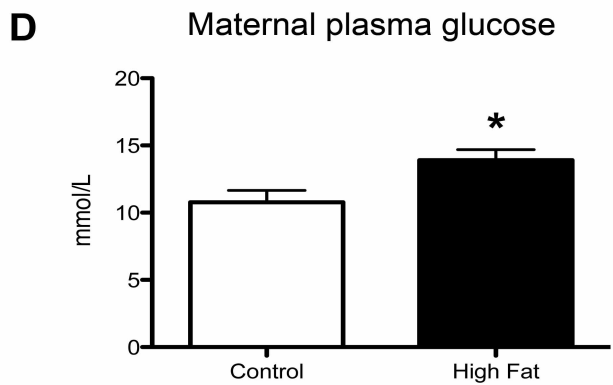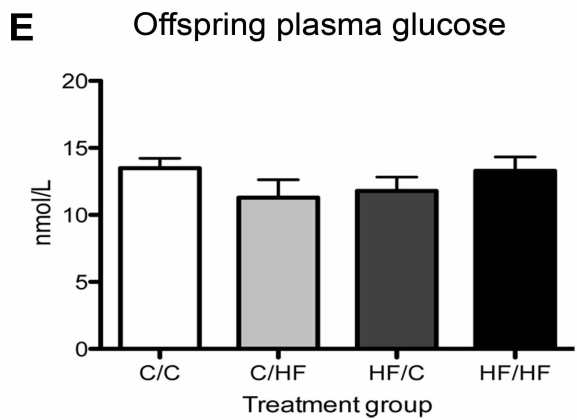
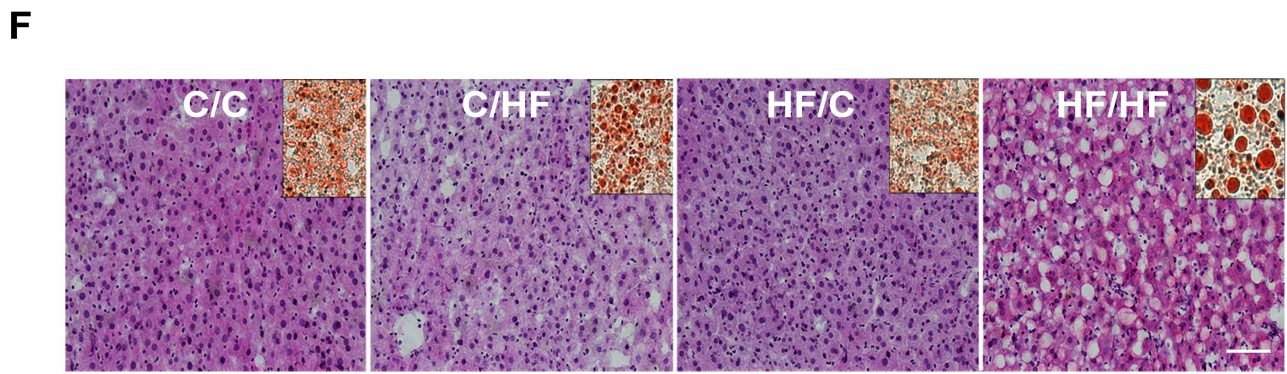

**Figures**

**Fig. 1** Model of pre- and postnatal high fat feeding. (**A**) Dams were maintained on either a standard (C) or high fat (HF) diet before conception and during gestation and lactation. At weaning, offspring were assigned either C or HF diet, generating four experimental groups: C/C, C/HF, HF/C, HF/HF, representing pre- and post-weaning diet, respectively. (**B**) Dams fed the HF diet weighed significantly more during both the preconception and gestational period than those on the C diet. *p<0.05 vs. control. (**C**) Offspring in the C/HF and HF/HF groups weighed significantly more than mice in the C/C and HF/C groups within 8 and 4-6 weeks post-weaning, respectively. *p<0.05 vs. C/C; #p<0.05 vs. C/HF; + p<0.05 vs. HF/C. (**D and E**) Plasma glucose levels were significantly higher in dams fed the HF diet compared to those fed the C diet (D), but did not differ between offspring dietary groups (E). *p<0.05 vs. control. (**F**) Liver H&E histology and Oil red O (insets) revealed normal tissue architecture in the C/C mice, mild steatosis in the C/HF group, a few lipid particles in the HF/C group and extensive fat accumulation in the HF/HF group. Data are presented as mean ± S.E.M. Scale bar = 200 μm.

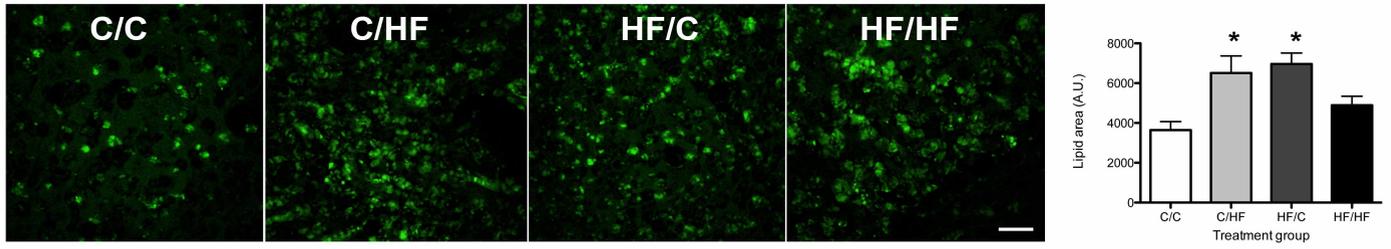
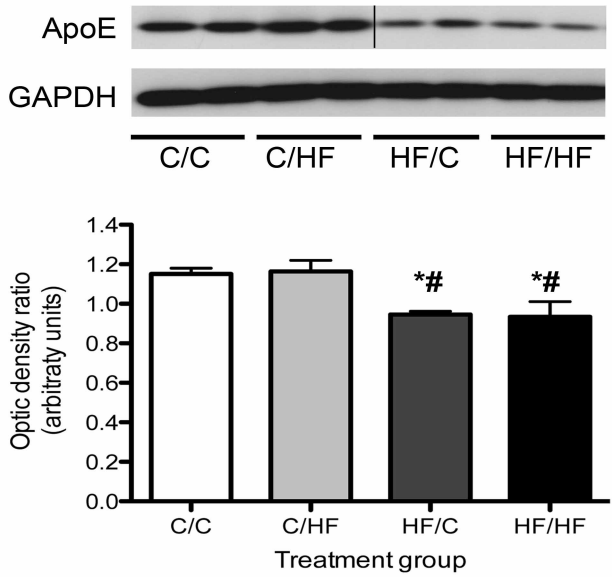
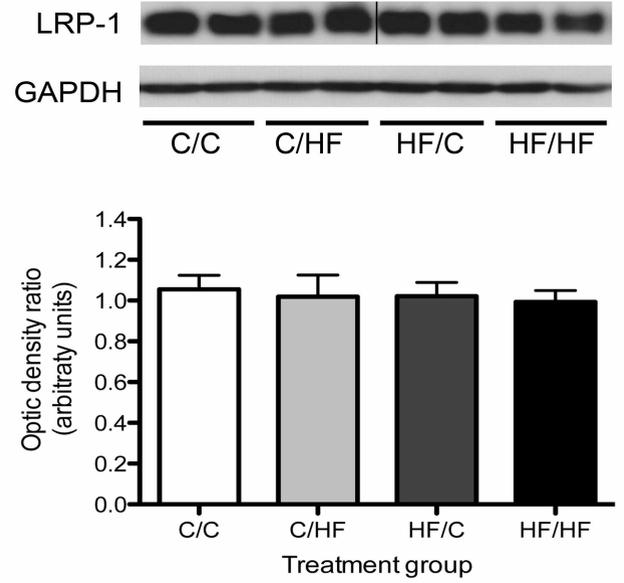
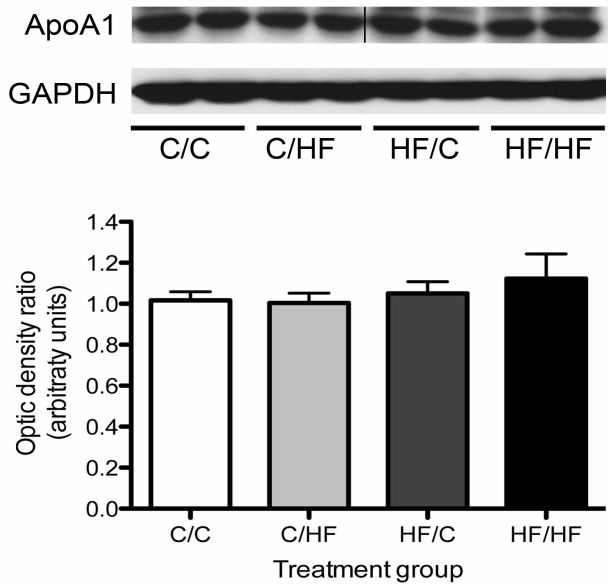
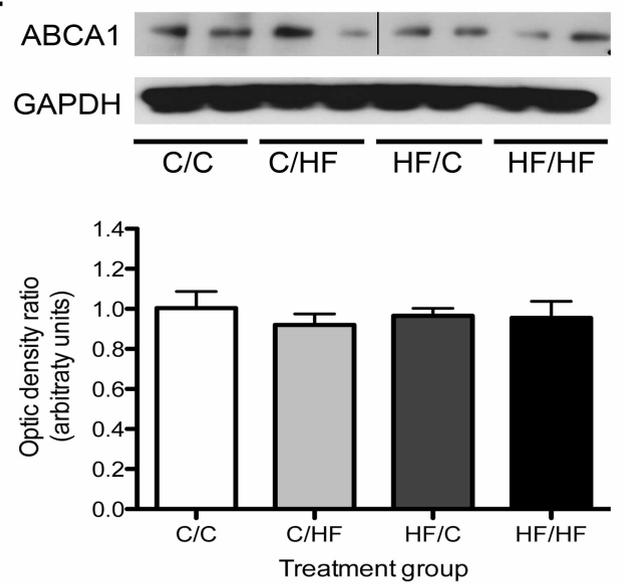

**Fig. 2** Effect of pre- and postnatal high fat diet on cerebral lipid droplet accumulation and markers of cholesterol transport in the hippocampus of adult offspring. (**A**) A significantly greater hippocampal area was covered by lipid droplets in the brains of C/HF and HF/C mice, but not HF/HF mice, compared to mice in the C/C group. *p<0.05 vs. C/C. (**B**) ApoE levels were significantly decreased in the brains of HF/C and HF/HF mice compared to both C/C and C/HF groups. *p<0.05 vs. C/C; #p<0.05 vs. C/HF. (**C-E**) Western blot analyses did not reveal any differences in the level of expression of low density lipoprotein-related receptor-1 (LRP-1, C), apolipoprotein A-I (ApoA-I, D) or ATP-binding cassette transporter (ABCA1, E) between offspring dietary groups. Data are presented as mean ± S.E.M. Scale bar = 20 μm. Gels were cropped to band of interest and black lines indicate non-contiguous lanes of bands run on the same gel.

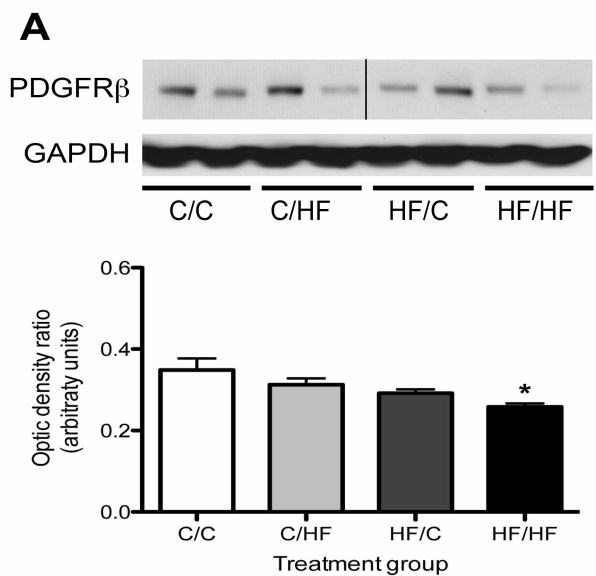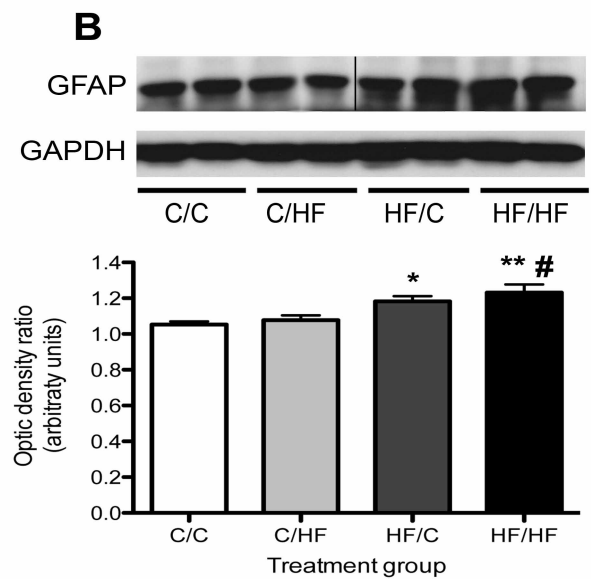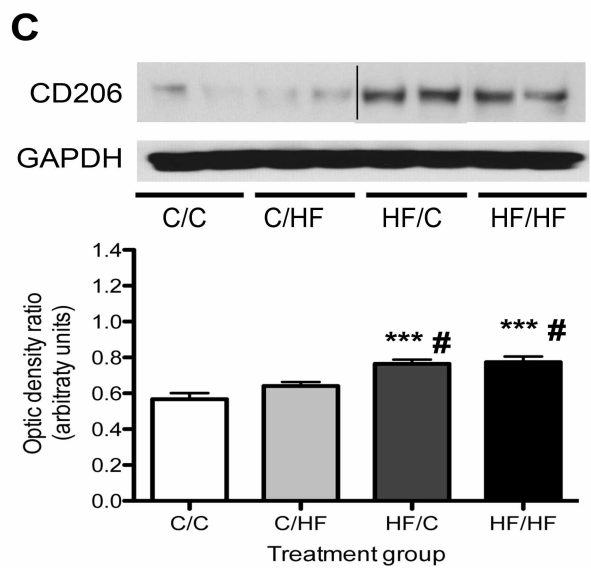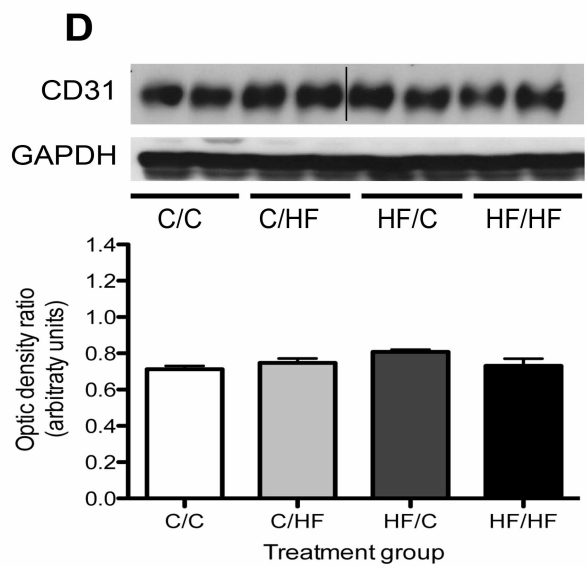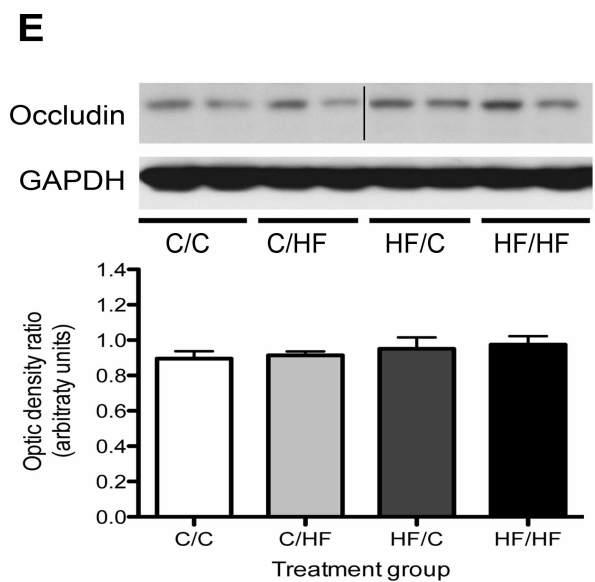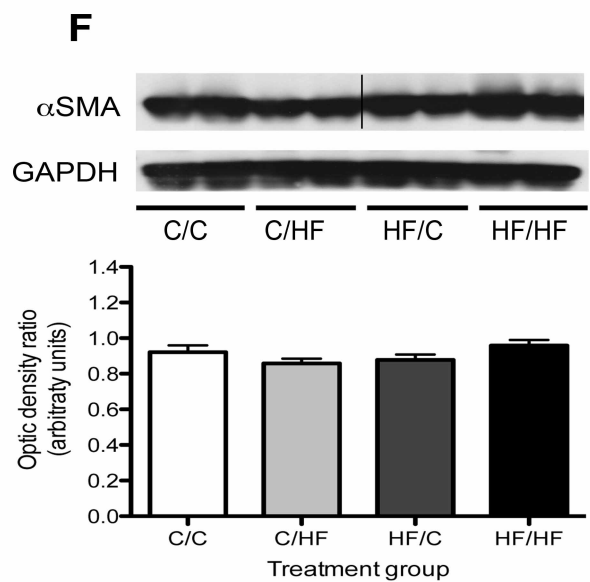

**Fig. 3** Effect of pre- and postnatal high fat diet on markers of neurovascular unit in the hippocampus of adult offspring. **(A)** Levels of platelet-derived growth factor receptor β (PDGFRβ), a marker of activated pericytes, were significantly decreased in the hippocampi of HF/HF mice compared to the C/C group. **(B)** Gial fibrillary acidic protein (GFAP) was significantly upregulated in both the HF/C and the HF/HF group versus the C/C group and between HF/HF and C/HF mice. **(C)** Levels of the perivascular macrophage CD206 receptor were significantly upregulated in the hippocampi of HF/C and HF/HF mice compared to C/C and C/HF groups. **(D-F)** No differences were observed in the levels of the endothelial cell marker CD31 (D), the tight junction protein occludin (E) or α-smooth muscle actin between any of the offspring groups (F). Data are presented as mean ± S.E.M. *p<0.05 vs. C/C; **p<0.01 vs. C/C; *** p<0.001 vs. C/C; #p<0.05 vs. C/HF. Gels were cropped to band of interest and black lines indicate non-contiguous lanes of bands run on the same gel.

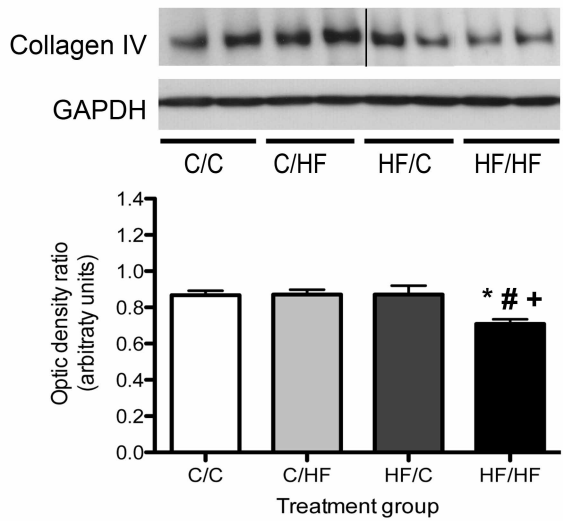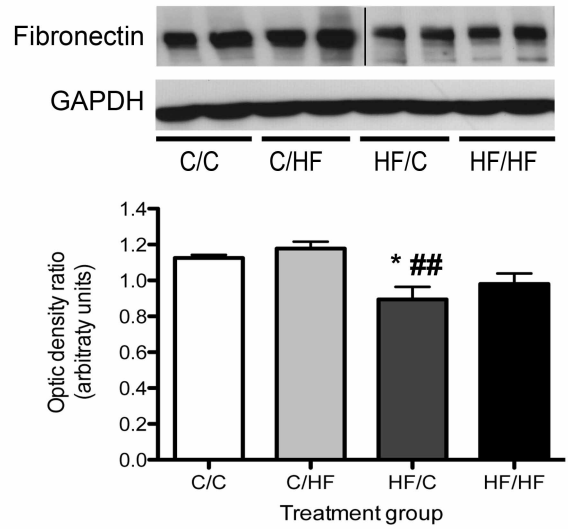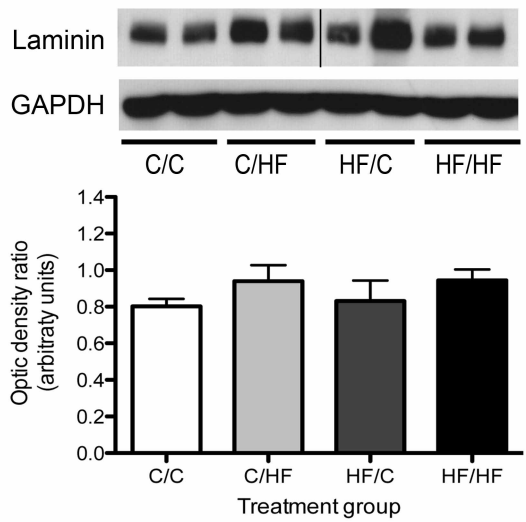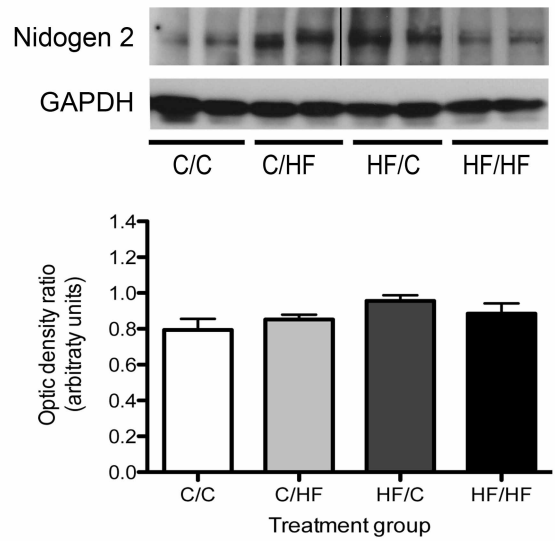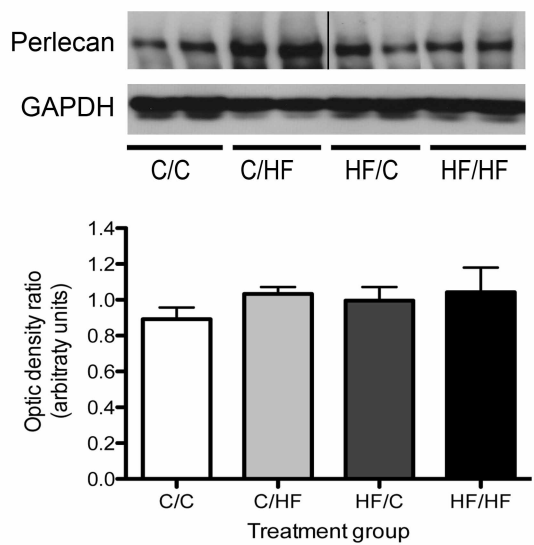

**Fig. 4** Assessment of pre- and postnatal high fat diet on cerebrovascular basement membrane proteins. **(A)** Levels of collagen IV were significantly decreased in HF/HF mice compared to all other dietary groups. **(B)** Fibronectin levels were significantly lower in the hippocampi of HF/C mice versus C/C and C/HF groups. **(C-E)** Levels of laminin (C), nidogen 2 (D) and perlecan (E) did not differ between offspring dietary groups. Data are presented as mean ± S.E.M. *p<0.05 vs. C/C; #p<0.05 vs. C/HF; ##p<0.01 vs. C/HF; + p<0.05 vs. HF/C. Gels were cropped to band of interest and black lines indicate non-contiguous lanes of bands run on the same gel.

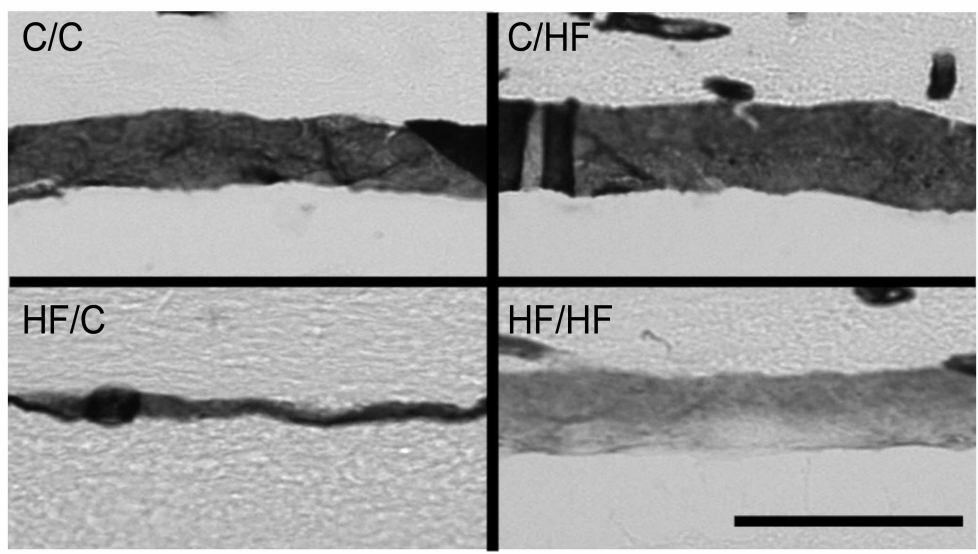

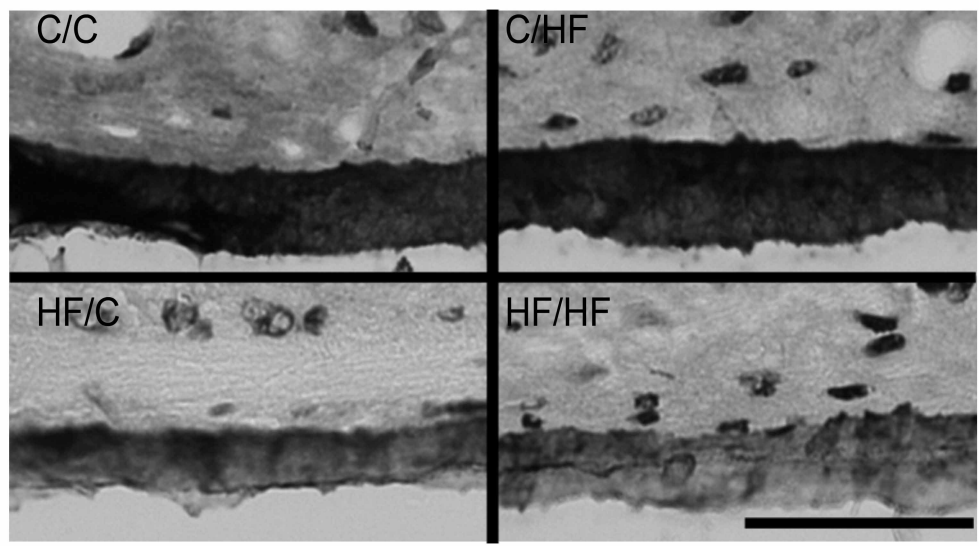

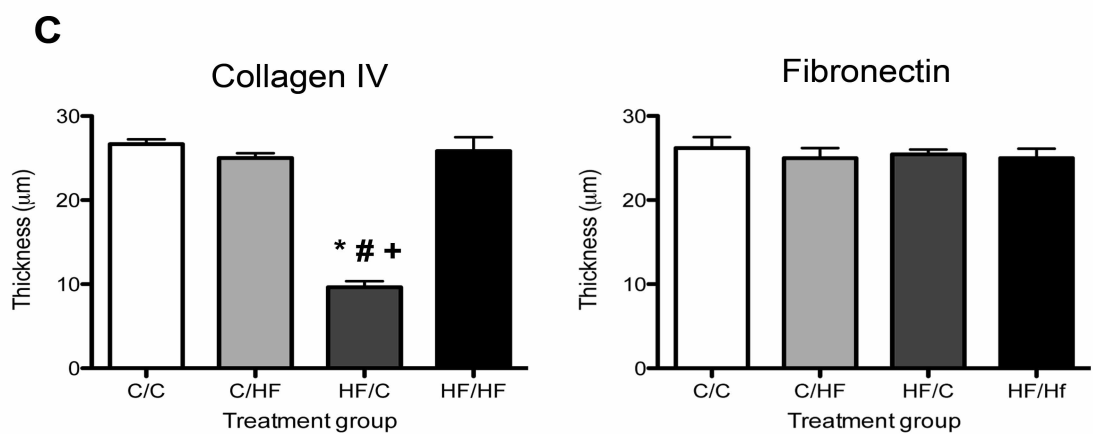

**Fig. 5** Effect of pre- and postnatal high fat diet on cerebrovascular morphology. **(A and B)** Morphologic analysis of collagen IV (A) and fibronectin (B) showed a smooth distribution along hippocampal arteries in the C/C group, with a slightly more rugged appearance in the C/HF mice. In the HF/C mice, vessels stained with collagen IV had a dramatically reduced diameter, while fibronectin expression appeared reduced compared to C/C and C/HF brains. Less staining of both collagen IV and fibronectin was also noted in the HF/HF brains. **(C)** Quantification of vessel diameter showed a significant reduction in the diameter of collagen IV-positive arteries in the HF/C group compared to all other dietary groups. The diameter of vessels stained with fibronectin did not vary between groups. Data are presented as mean ± S.E.M. *$p<0.05$ vs. C/C, #$p<0.05$ vs. C/HF, +$p<0.05$ vs. HF/HF. Scale bars = 50 μm.

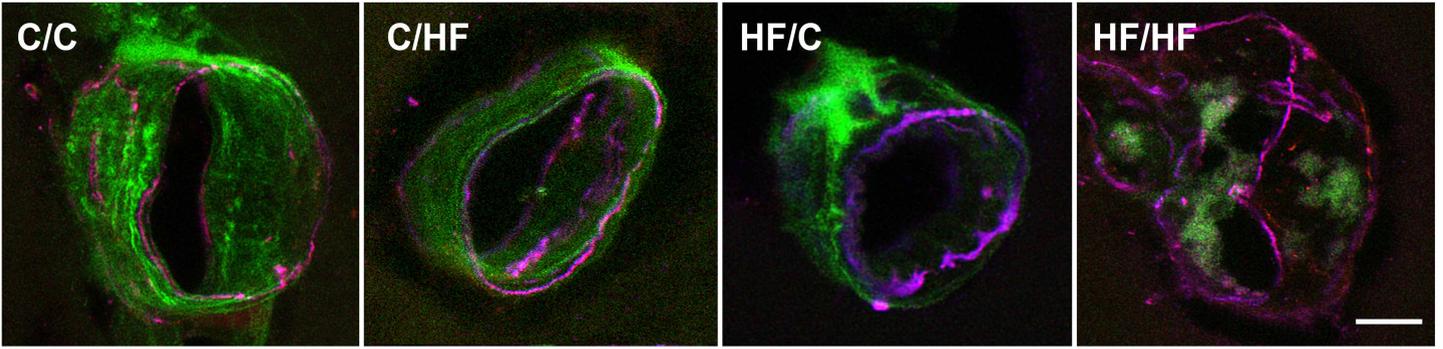
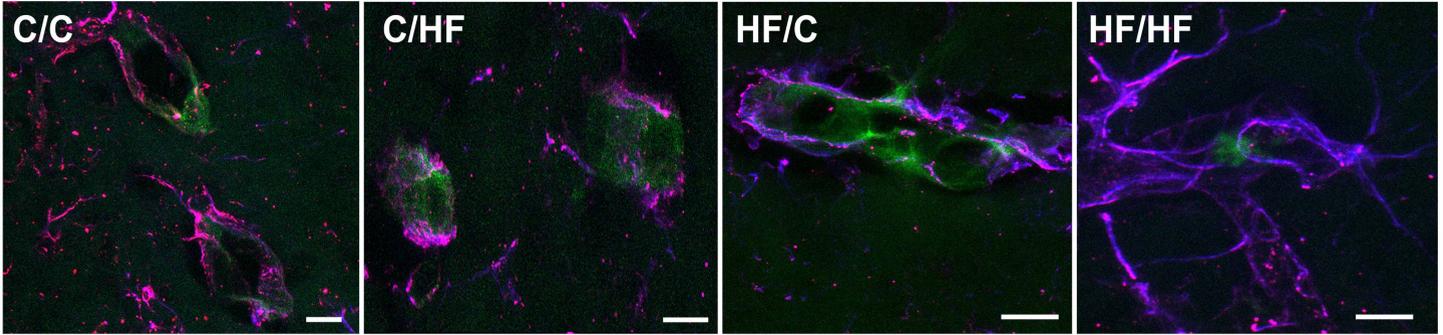
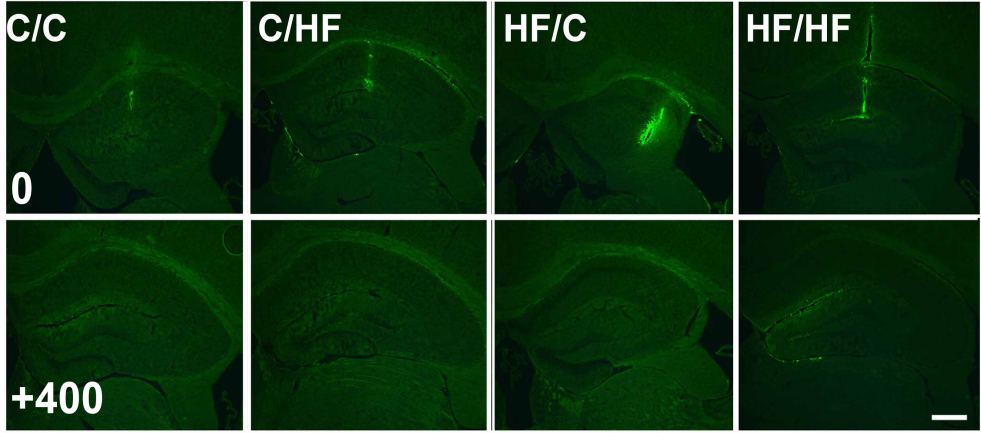
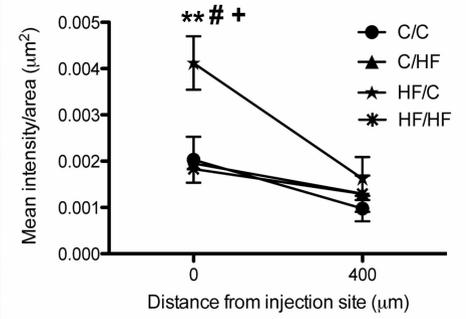

**Fig. 6** Perivascular drainage of Aβ in the brains of offspring fed a pre- or postnatal high fat diet. (**A**) HyLite Fluor 488-conjugated human Aβ40 (green) injected into the hippocampus of C/C and C/HF mice distributed in an even pattern along laminin-positive (red) basement membranes in hippocampal arteries (identified by α smooth muscle actin staining, blue). Colocalization of laminin and α smooth muscle actin appears pink. In HF/C mice, Aβ drainage along the basement membrane appeared ruffled and disrupted compared to the C/C group. Within the arteries of HF/HF mice, Aβ was observed as large deposits that did not appear to drain. (**B**) Double labeling immunocytochemistry of hippocampal vessels with antibodies against GFAP (blue) and ApoE (red) following intracerebral injection of Aβ40 (green) showed a decrease in the expression of ApoE in the brains of HF/C and HF/HF mice compared to C/C and C/HF mice. Colocalization of GFAP and ApoE at the blood vessels appears pink. GFAP and ApoE localized to perivascular astrocytes, but did not colocalize with Aβ in any of the treatment groups. (**C**) The mean fluorescence intensity of Aβ at the site of injection (0) was significantly higher in the hippocampi of HF/C mice compared to the other dietary groups, but did not differ at 400 μm (+400) away from the site of injection. Data are presented as mean ± S.E.M. **$p<0.01$ vs. C/C; #$p<0.01$ vs. C/HF; +$p<0.01$ vs. HF/HF. Scale bars: A = 7.5 μm; B = 10 μm; C = 400 μm.

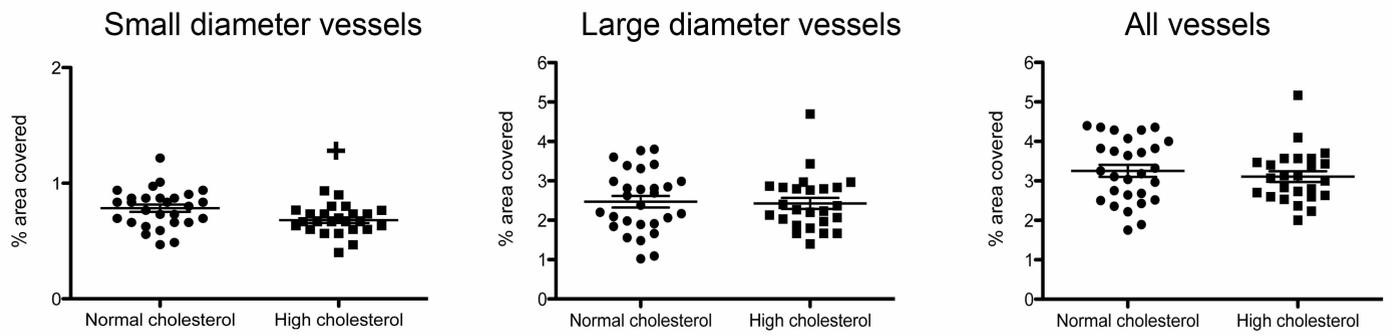
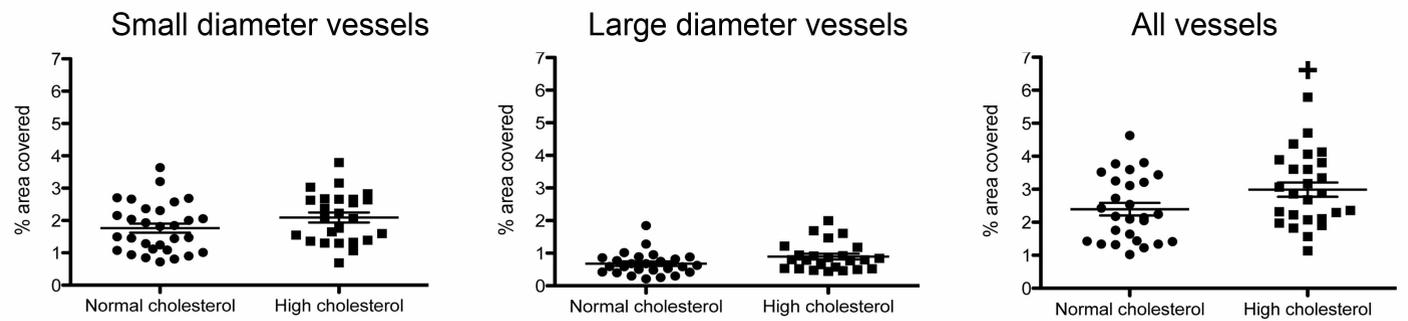
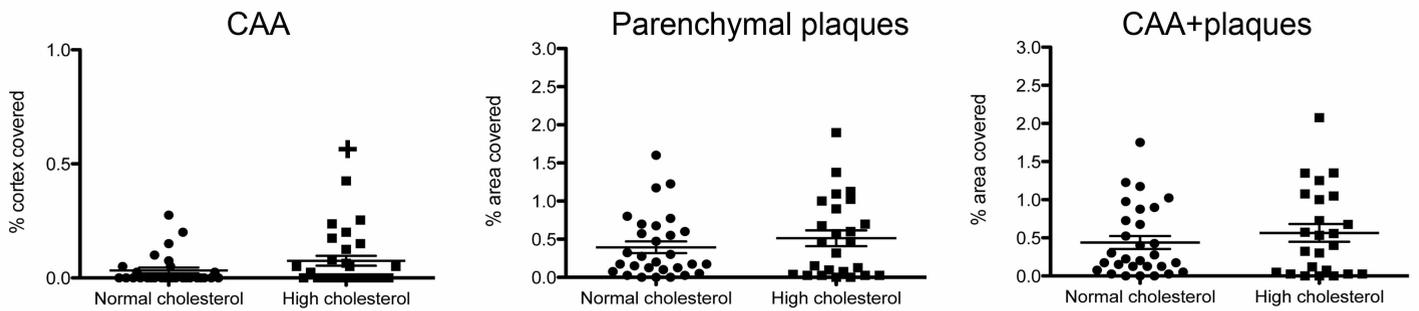
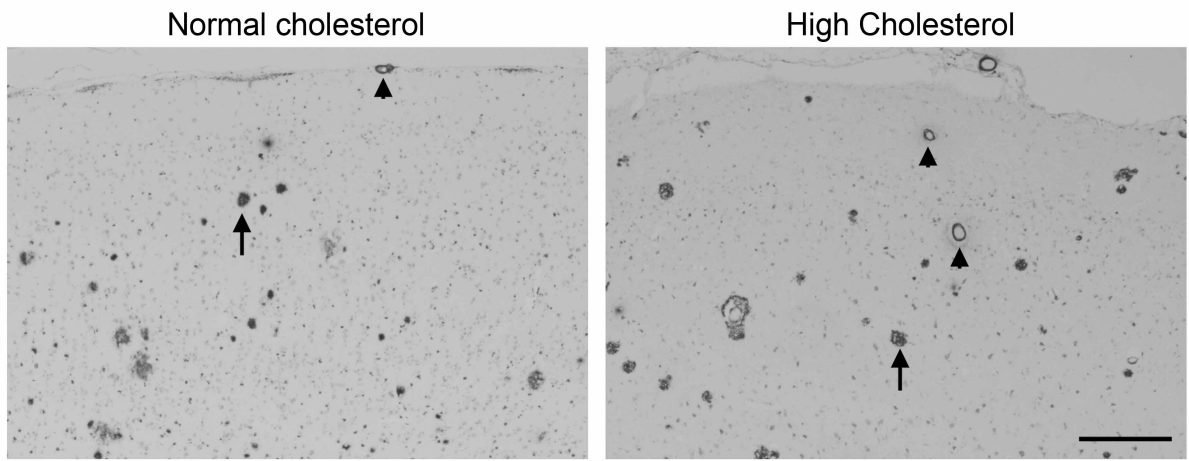

**Fig. 7** Effect of hypercholesterolemia on basement membrane protein expression and Aβ load in human brains.  (**A**) Tissue sections from the occipital cortex of individuals with normal and high levels of cholesterol were assessed by immunocytochemistry for the expression of collagen IV.  A significant reduction in cortical coverage by small diameter, collagen IV-positive blood vessels was observed in individuals with hypercholesterolemia.  No differences were noted between groups in the percent cortical coverage by large diameter vessels or in total collagen IV-positive vessels.  (**B**) Fibronectin expression did not differ between normal and high cholesterol groups when small and large diameter vessels were analyzed separately, but was significantly upregulated in the high cholesterol group when the cortical coverage of fibronectin staining for all vessels was calculated.   (**C**) Brains from humans with high levels of cholesterol had a significantly greater load of cerebral amyloid angiopathy (CAA) than those with normal cholesterol.  Cholesterol levels did not impact upon the percent of cortical coverage by parenchymal plaques or total Aβ deposition.  (**D**) Micrographs of brain tissue sections stained for Aβ-positive plaques (arrows) and CAA (arrow heads), showing greater CAA load in individuals with high levels of cholesterol.  Data are presented as mean ± S.E.M. +$p<0.05$ vs. normal cholesterol.  Scale bar = 400 μm.

**Table 1** Fatty acid species of phosphatidylcholine (PC) and phosphatidylethanolamine (PE) that differ between dietary groups

|          | C/C          | C/HF               | HF/C         | HF/HF             |
|----------|--------------|--------------------|--------------|-------------------|
| PC 38:6  | 4.23 ± 0.06  | 4.61 ± 0.06$^{a,d}$ | 4.31 ± 0.01  | 4.23 ± 0.11       |
| PC 38:5  | 2.32 ± 0.04  | 2.33 ± 0.03        | 2.42 ± 0.05  | 2.19 ± 0.05$^c$   |
| PC 38:4  | 6.71 ± 0.09  | 7.03 ± 0.08$^d$    | 6.97 ± 0.09  | 6.62 ± 0.09       |
| PC 40:6  | 1.75 ± 0.02  | 2.00 ± 0.05$^{a,d}$ | 1.89 ± 0.03  | 1.81 ± 0.05       |
| PE 38a:0 | 1.30 ± 0.05  | 1.10 ± 0.07        | 1.16 ± 0.07  | 1.34 ± 0.07$^{b,c}$ |
| PE 38:4  | 12.56 ± 0.10 | 12.51 ± 0.21       | 13.00 ± 0.30 | 11.58 ± 0.20$^c$  |
| PE 40:3  | 2.52 ± 0.08  | 2.44 ± 0.13        | 2.25 ± 0.15  | 2.96 ± 0.14$^c$   |
| PE 42a:2 | 11.33 ± 0.45 | 11.23 ± 0.50       | 11.01 ± 0.55 | 13.16 ± 0.48$^c$  |

$^a$p<0.05 vs. C/C; $^b$p<0.05 vs. C/HF; $^c$p<0.05 vs. HF/C; $^d$p<0.05 vs. HF/HF. Values represent mean percentage of total phospholipid content ± S.E.M.

**Table 2** Summary of changes to neurovascular unit proteins between diet groups

| Comparison | C/HF | HF/C | HF/HF |
|---|---|---|---|
| **C/C vs.** | — | ↑ astrocytes | ↑ astrocytes |
|  | — | ↑ perivascular macrophages | ↑ perivascular macrophages |
|  | — | ↓ ApoE | ↓ ApoE |
|  | — | ↓ fibronectin | ↓ collagen IV |
|  |  |  | ↓ pericytes |
|  |  |  |  |
| **C/HF vs.** | — | ↑ perivascular macrophages | ↑ perivascular macrophages |
|  | — | ↓ ApoE | ↓ ApoE |
|  |  | ↓ fibronectin |  |
|  |  |  |  |
| **HF/C vs.** | — | — | ↓ collagen IV |

**Table 3** List of primary antibodies used for Western blotting

| Antibody | Dilution | Source |
| --- | --- | --- |
| α-Smooth muscle actin | 1:500 | Sigma-Aldrich, Dorset, UK |
| Apolipoprotein A-I (ApoA-I) | 1:1500 | Merck Millipore, Nottingham, UK |
| Apolipoprotein E (ApoE) | 1:750 | Merck Millipore, Nottingham, UK |
| ATP-binding cassette transporter A1 (ABCA1) | 1:700 | Thermo Scientific, Loughborough, UK |
| CD31 | 1:500 | AbDSerotec, Kidlington, UK |
| CD206 | 1:400 | AbDSerotec, Kidlington, UK |
| Collagen IV | 1:500 | Abcam, Cambridge, UK |
| Fibronectin | 1:1000 | AbDSerotec, Kidlington, UK |
| Glial fibrillary protein (GFAP) | 1:1000 | Dako, Cambridgeshire, UK |
| Glyceraldehyde 3-phosphate dehydrogenase (GAPDH) | 1:50,000 | Sigma-Aldrich, Dorset, UK |
| Laminin | 1:500 | Sigma-Aldrich, Dorset, UK |
| Low density receptor-related protein 1 (LRP-1) | 1:750 | Insight Biotechnology, Middlesex, UK |
| Nidogen 2 | 1:1000 | Sigma-Aldrich, Dorset, UK |
| Occludin | 1:500 | Life Technologies, Paisley, UK |
| Perlecan | 1:500 | Merck Millipore, Nottingham, UK |
| Platelet-derived growth factor receptor-β (PDGFRβ) | 1:500 | Abcam, Cambridge, UK |